%% file: High_Power_Piezo_Driver.tex
\documentclass[journal]{IEEEtran}
\ifCLASSINFOpdf
\else
\fi

\newcommand{\PreserveBackslash}[1]{\let\temp=\\#1\let\\=\temp}
\newcommand{\TNSJournal}[1]{\def\@TNSJournal{#1}}

\usepackage{color}
\usepackage{array}
\usepackage{multirow}
\usepackage{textcomp}  
\usepackage{cite}
\usepackage{graphicx}
\usepackage{cite}      
\usepackage{subfigure} 
\usepackage[T1]{fontenc}
\usepackage{siunitx}
\usepackage{fixltx2e} 

\DeclareSIQualifier\peaktopeak{pp}

\graphicspath{{./fig/}}

\begin{document}

\title{High-power Piezoelectric Tuner Driver for Lorentz Force Compensation}

\author{
	D.~Makowski,~\IEEEmembership{Member,~IEEE}, A.~Mielczarek, P.~Perek, A.~Szubert, P.~Plewi\'nski, G.~Jab\l{}o\'nski, W.~Cichalewski, A.~Napieralski,~\IEEEmembership{Senior Member,~IEEE}

	\thanks{Manuscript received June 15, 2018; January 11, 2018.}
	
	\thanks{D.~Makowski, A.~Mielczarek, P.~Perek, A.~Szubert, P.~Plewi\'nski, G.~Jab\l{}o\'nski, W.~Cichalewski, A.~Napieralski,
			are with the Lodz University of Technology, Poland (e-mail: dmakow@dmcs.pl)}
}

\maketitle
\thispagestyle{empty}

\input{src/0_Abstract}
\input{src/Extended_Abstract}

\ifx\TNSJournal\undefined
\IEEEtriggeratref{29}
\else
\IEEEtriggeratref{29}
\fi

\bibliographystyle{IEEEtran}
\bibliography{../bib/DM_Habilitation_2016.04.02,../bib/rt10,../bib/rt12,../bib/rt14,../bib/ATCA_Papers_2008.10.08,../bib/dmakow_2010.05.15,../bib/PhD_2010.05.15,../bib/ieee_standards,../bib/ijmcs,../bib/mixdes,../bib/IAS,../bib/ieee1588,../bib/NDS}

\end{document}

%% file: src/0_Abstract.tex
\begin{abstract}
Superconducting Radio Frequency (SRF) cavities are used in modern accelerators to efficiently accelerate particles.
When cavity is supplied with pulsed RF field it undergoes a mechanical strain due to the Lorentz force.
The resulting deformation causes dynamic detuning of which magnitude depends on mechanical properties of the cavity,
RF pulse rate and their profile. This effect causes considerable loss of acceleration performance.
Therefore, it is usually actively compensated, most commonly with fast piezoelectric actuators.

MicroTCA.4 standard was developed to accommodate control and data acquisition electronic systems of large-scale physics applications. The paper presents the design of a high-power amplifier implemented using the MicroTCA.4 technology. The design of the driver was optimized for driving large-capacitance piezo actuators. Several possible architectures of the driver are presented and compared, taking into consideration the power and cooling limitations of MicroTCA.4. The design of a two-channel piezo driver and its initial laboratory test results are also discussed.

\end{abstract}

\begin{IEEEkeywords}
Lorentz Force Compensation, Piezoelectric Tuner, Pulse-Width Modulation, Micro Telecommunications Computing Architecture, Linear Accelerator
\end{IEEEkeywords}

%% file: src/Extended_Abstract.tex
\section{Introduction}

Many control and data acquisition systems used in large scale physics experiments are developed with MicroTCA.4 (Micro Telecommunications Computing Architecture) standard. Good examples include the Low-Level Radio Frequency (LLRF) control systems of linear accelerators such as the Free-Electron Laser in Hamburg (FLASH), European X-ray Free-Electron Laser (E-XFEL) or European Spallation Source (ESS) or diagnostics systems of the ITER tokamak. 

Application of modern standards, such as MicroTCA, supporting the hot-swap technology, power supply, cooling and data transmission fabric redundancy as well as intelligent chassis management allow to design scalable high-reliability control and telecommunication systems~\cite{makowski_method_2016}. 

In order to maintain high acceleration efficiency the superconducting cavities require dynamic resonance frequency tuning. This is commonly done with piezoelectric actuators. It is desired to integrate the Lorentz Force Detuning (LFD) compensation system together with the LLRF control hardware. Therefore, it is convenient to design the piezo driver as Advanced Mezzanine Card (AMC) or Rear Transition Module (RTM) that is installed in the same MicroTCA.4 chassis. 

\section{Limitations of MicroTCA.4 standard} 

The MicroTCA.4 standard, developed in 2011~\cite{picmg_mtca.4_2011}, is based on the original PCI Industrial Computer Manufacturers Group (PICMG) MicroTCA.0 specification~\cite{picmg_mtca.0_2006} and therefore inherits the main limitations concerning power consumption and heat generation in a single slot. The MicroTCA.4 standard allows dissipating up to \SI{80}{\watt} in a single AMC slot. When the Rear Transition Module is used the power provided by the MicroTCA power supply is shared between the AMC (\SI{50}{\watt}) and the RTM device (\SI{30}{\watt}). In such situation the cooling capabilities are limited to \SI{80}{\watt} for AMC and \SI{30}{\watt} for RTM. 

\begin{figure}[htb]
\centering
\includegraphics[width=0.5\textwidth]{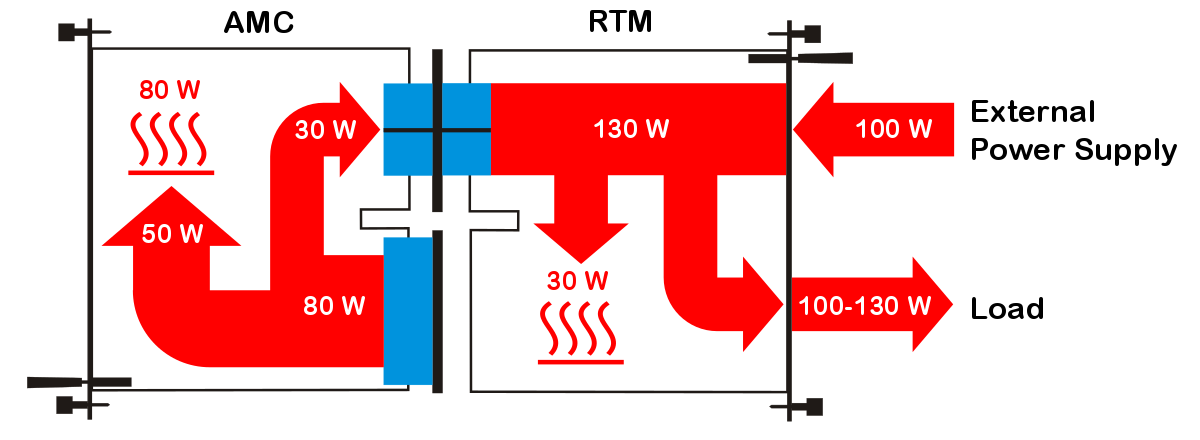}
\caption{Power distribution for AMC and RTM for high power application}
\label{fig:AMC_RTM_power}
\end{figure}

The hard limits of MicroTCA.4 technology make designing a high power piezo driver a nontrivial task. A few scenarios were carefully analyzed, including driver implemented as AMC or RTM cards, to find the optimal solution. Eventually the architecture involving an external power supply was selected. The power and cooling capabilities for \SI{80}{\watt} internal and \SI{100}{\watt} external power supplies are illustrated in Figure~\ref{fig:AMC_RTM_power}. 

The external power supply can provide much more power than available from MicroTCA internal power supply. However, the cooling is still limited to \SI{30}{\watt} on the RTM side. Therefore, the design of a large-power piezo driver suitable for accelerators operating with long RF pulses, such as the ESS machine built in Sweden, requires a high-efficiency amplifier.

\section{High-power piezo actuator module}

Piezo actuators manufactured by Noliac or Physik Instrumente, which are commonly used for LFD compensation, require driving voltages up to \SI{200}{\volt\peaktopeak} (\SI{\pm100}{\volt}) at cryogenic temperatures. The multilayer piezo plate actuator stack capacitance depends on its blocking force and the number of cells. For example, in Noliac NCE51F devices it can reach \SI{64}{\micro\farad} at room temperature and \SI{21}{\micro\farad} in cryogenic conditions. Usually, two redundant amplifier channels are implemented in piezo tuning system for a single cavity. The channels can also be used in parallel to increase the blocking force and therefore allow higher force compensation. 

Simple LFD compensation systems use a pulsed sinusoidal \SI{1}{\kilo\hertz} repetitive signal for driving the piezo actuators. For short RF pulses a single period of sine wave just before the RF pulse is able to compensate the Lorentz force~\cite{PrzygodaTNS09}. For longer RF pulses the driver should be able to provide more pulses. More advanced tuning algorithms with cavity detuning identification could require more sophisticated signals with components of frequencies reaching tens of \SI{}{\kilo\hertz}.

\begin{figure}[htb]
\centering
\includegraphics[width=0.3\textwidth]{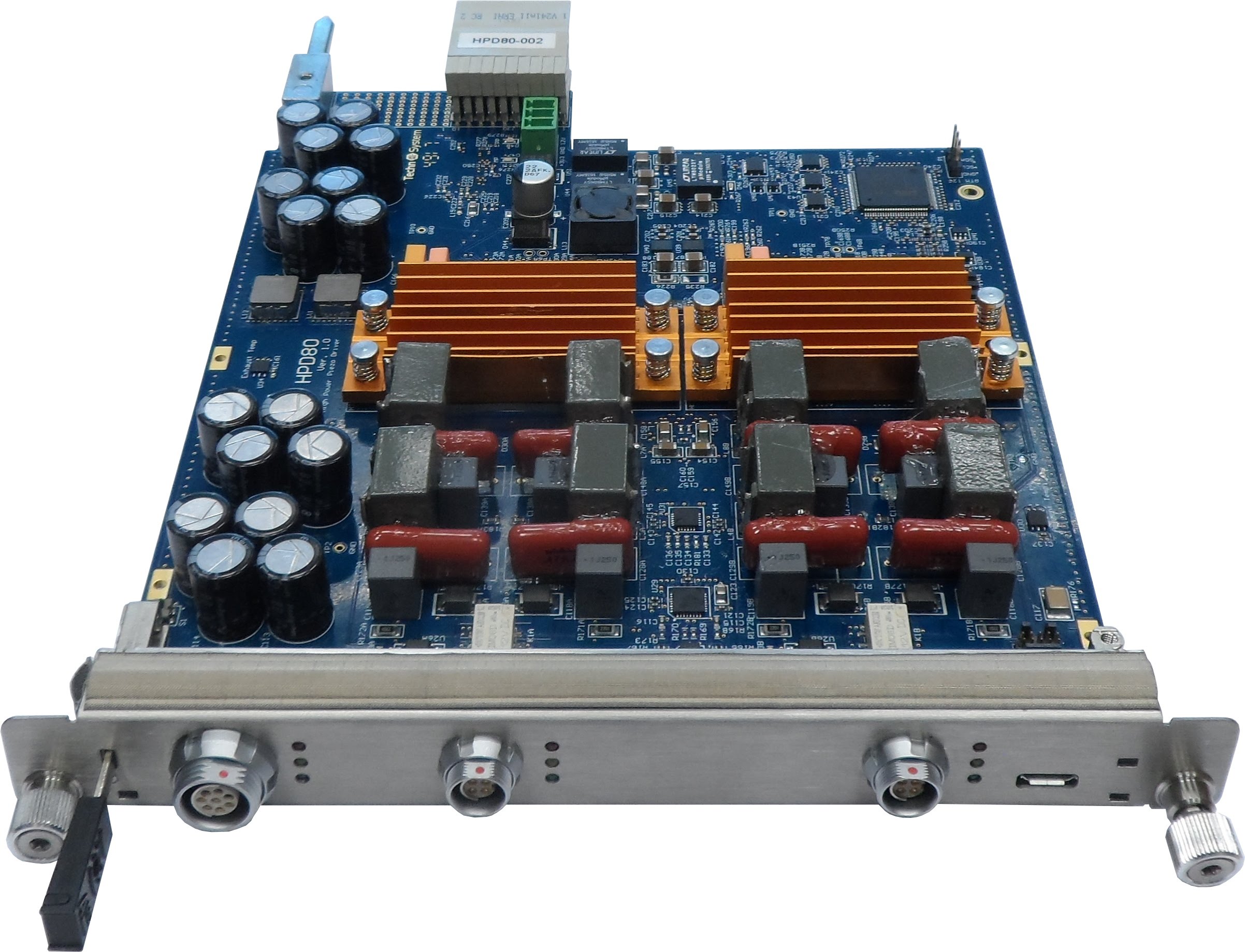}
\caption{A high power piezo driver implemented as Rear Transition Module}
\label{fig:hpd80}
\end{figure}

A high power 2-channel piezo driver implemented as an RTM module was developed and tested. The piezo driver module is presented in Figure~\ref{fig:hpd80}.

In order to overcome the \SI{30}{\watt} cooling limitations on RTM side a high-efficiency power amplifier is needed. A typical linear Class-AB amplifier is clearly not suitable for such applications, due to its poor energy efficiency limited to 50\%. In case of a single channel \SI{100}{\watt} driver the AB amplifier would have to be able to dissipate at least \SI{50}{\watt}.

Therefore, a high-efficiency Class-D amplifier based on a Pulse-Width Modulation (PWM) controller from International Rectifier was used as a basis of the MicroTCA.4 driver. The solution is equipped with two \SI{100}{\watt} channels and it is suitable for driving large capacitance piezo actuators operating in pulsed or even continuous mode. The driver uses an external \SI{200}{\watt} \SI{\pm50}{\volt} power supply and therefore allows controlling the piezo with amplitude reaching \SI{200}{\volt\peaktopeak}. The power generated by the high power section of the device is expected not to exceed \SI{20}{\watt}.
Both channels of the driver are configurable and can operate in either driver or sensor mode. The device is equipped with driver protection circuitry and advanced health monitoring based on the Intelligent Platform Management Interface (IPMI) standard. The solution also offers built-in protection mechanisms that monitor the control signals and protect the actuator. 

\section{Initial results}

The RTM piezo driver is equipped with control DACs which are connected to the Zone 3 D1.1 class connectors. The solution needs an AMC module compatible with MicroTCA.4 and equipped with an FPGA device that can generate control signals driving the piezo actuator.

The tests of the driver have been performed in a 7-slot, 5~U, 42~HP Schroff MicroTCA.4 crate equipped with NAT PHYS~80 Module Carrier Hub and Kontron AM4020 CPU. 
The driver RTM has been hosted by the MFMC AMC module with Artix-7 FPGA device. For the tests, a dedicated FPGA firmware implementing an arbitrary waveform generator has been developed. It enabled controlling the ADC circuits on the RTM from the CPU via the PCI Express bus with help of a dedicated driver and a C++ application using the Qt library. The output of the driver was first evaluated with various film capacitors ranging from \SI{2.2}{\micro\farad} up to \SI{50}{\micro\farad}.

\begin{figure}[htb]
\centering
\includegraphics[width=0.3\textwidth]{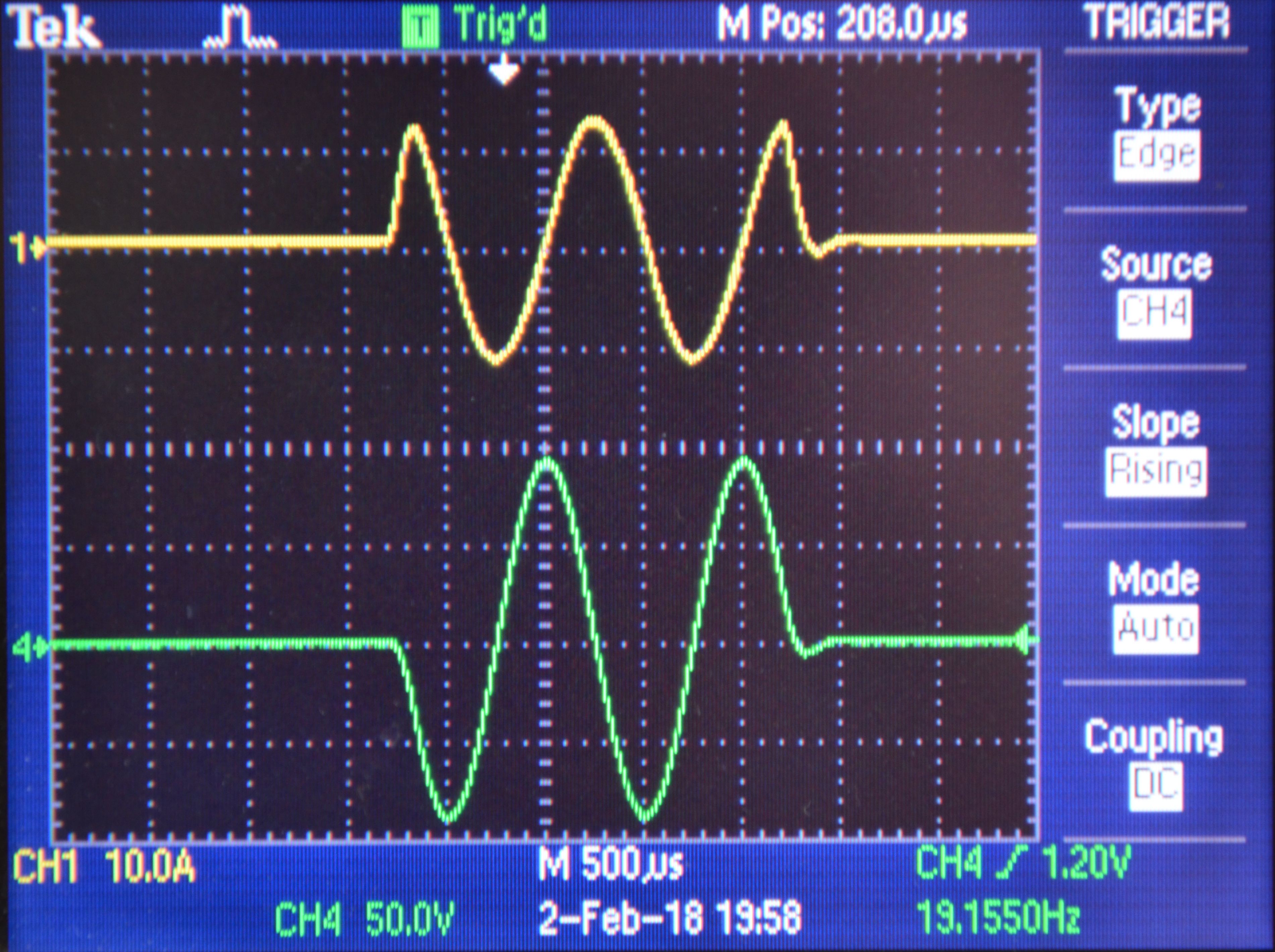}
\caption{The example load current and voltage for the \SI{20}{\micro\farad} capacitive load.}
\label{fig:2pulses}
\end{figure}

Next, the driver was applied to excite two Noliac NAC 2022 H30 piezo actuators operated in room temperature with measured capacitance of \SI{6.6}{\micro\farad}. The voltage waveform consisting of several periods of 1~kHz sine wave with \SI{14}{\hertz} repetition rate has been applied. For the \SI{20}{\micro\farad} capacitive load the driver was able to provide 10 pulses of \SI{1}{\kilo\hertz} sine wave with the amplitude of \SI{180}{\volt\peaktopeak} without overheating. Figure \ref{fig:2pulses} presents example of load current and voltage measurements performed during the tests. 

\section{Conclusion}
We developed the first high-power piezo driver for MicroTCA.4-based systems that can control the largest piezoelectric actuators with the capacitances over \SI{20}{\micro\farad}. The driver fulfills all the MicroTCA requirements, in particular the RTM dissipated power limit of \SI{30}{\watt}.
The initial tests have shown, that the RTM can interoperate with the AMC module and can drive loads up to 
\SI{50}{\micro\farad}. We are planning further tests in cavity assemblies in cryostats and accelerators.

The current, high power version of the driver requires an external power supply. However, if the drive requirements were relaxed, it would also be possible to build a version supplied solely from an AMC module.